\documentclass[aps,prl,twocolumn,superscriptaddress]{revtex4-1}
\usepackage{graphicx}

\begin{document}

\title{Multi-Spacecraft Measurement of Turbulence within a Magnetic Reconnection Jet}

\author{K. T. Osman}
\email{k.t.osman@warwick.ac.uk}
\affiliation{Centre for Fusion, Space and Astrophysics; University of Warwick, Coventry, CV4 7AL, United Kingdom}

\author{K. H. Kiyani}
\affiliation{Laboratoire de Physique des Plasmas, \'Ecole Polytechnique, Route de Saclay, 91128 Palaiseau, France}
\affiliation{Centre for Fusion, Space and Astrophysics; University of Warwick, Coventry, CV4 7AL, United Kingdom}

\author{W. H. Matthaeus}
\affiliation{Bartol Research Institute, Department of Physics and Astronomy, University of Delaware, Delaware 19716, USA}

\author{B. Hnat}
\affiliation{Centre for Fusion, Space and Astrophysics; University of Warwick, Coventry, CV4 7AL, United Kingdom}

\author{S. C. Chapman}
\affiliation{Centre for Fusion, Space and Astrophysics; University of Warwick, Coventry, CV4 7AL, United Kingdom}
\affiliation{Department of Mathematics and Statistics, University of Troms\o, N-9037 Troms\o, Norway}

\author{Yu. V. Khotyaintsev}
\affiliation{Swedish Institute of Space Physics, Uppsala, Sweden}

\date{\today}

\begin{abstract}

The relationship between magnetic reconnection and plasma turbulence is investigated using multipoint \textit{in-situ} measurements from the Cluster spacecraft within a high-speed reconnection jet in the terrestrial magnetotail. We show explicitly
that work done by electromagnetic fields on the particles, $\mathbf{J}\cdot\mathbf{E}$, has a non-Gaussian distribution and is concentrated in regions of high electric current density. Hence, magnetic energy is converted to kinetic energy in an intermittent manner. Furthermore, we find the higher-order statistics of magnetic field fluctuations generated by reconnection are characterized by multifractal scaling on magnetofluid scales and non-Gaussian global scale invariance on kinetic scales. These observations suggest $\mathbf{J}\cdot\mathbf{E}$ within the reconnection jet has an analogue in fluid-like turbulence theory in that it proceeds via coherent structures generated by an intermittent cascade. This supports the hypothesis that turbulent dissipation is highly nonuniform, and thus these results could have far reaching implications for space and astrophysical plasmas.

\end{abstract}


\maketitle

\textit{Introduction}.---Turbulence is a universal nonlinear phenomenon that is ubiquitous in space plasmas \citep{ZimbardoEA2010}. It produces a cascade of coherent structures in neutral fluids \citep{AnselmetEA1984} and plasmas \citep{MatthaeusMontgomery1980,KarimabadiEA13}. These are concentrated structures that are phase correlated over their spatial extent and have relatively long lifetimes, such as current or vorticity sheets. Indeed, current sheets have been observed extensively 
in turbulent plasmas, and are associated with magnetic reconnection in the solar wind \citep{OsmanEA2014}, at the magnetopause
\citep{MozerEA2002}, and in the magnetosheath \citep{RetinoEA2007,SundkvistEA07}. 
In the terrestrial plasma sheet, \textit{in-situ} coherent structures display signatures of intermittent turbulence \citep{VorosEA2004,WeygandEA2005} in the form of rare large amplitude fluctuations that are highly non-Gaussian. These spatially inhomogeneous turbulent flows have been proposed as central to plasma sheet dynamics \citep{BorovskyEA1997,BorovskyFunsten2003,Chang1999}, and thus critical to understanding how stored electromagnetic energy in the magnetotail is converted into plasma energy. Here we consider whether coherent structures generated by magnetic reconnection reflect the nonlinear dynamics of intermittent turbulence and might be sites of nonuniform dissipation. These longstanding questions \citep{MatthaeusLamkin1986} are the subject of this Letter. 

The nature of turbulent dissipation within collisionless plasmas remains an open problem. A strong turbulent cascade is far from equilibrium and smaller scale behavior is driven by larger scale dynamics, with faster response times for decreasing scales \citep{MatthaeusEA2014}. In addition, intermittent turbulence generates small-scale coherent structures that are responsible for nonuniform dissipation \citep{Frisch1995,Biskamp2003}. For neutral fluids, the Kolmogorov refined similarity hypothesis (hereafter KRSH; \citep{Kolmogorov1962,Obukhov1962}) relates 
the statistics of increments of the velocity field 
on a given spatial scale 
to local averages of the dissipation rate on the same scale. 
Hence, large intermittent fluctuations on small scales are 
concomitant with high local concentrations of dissipation. 
While KRSH is unproven, it is well supported in hydrodynamics \citep{SreenivasanAntonia1997} and lies at the heart of modern fluid turbulence theory. In contrast, KRSH lacks verification and even precise formulation for collisionless plasmas since these introduce significant complications that have not yet been overcome. 
Amongst these complications is the inability to write an explicit form
of the dissipation function, which is well known in viscous
hydrodynamics and visco-resistive magnetohydrodynamics.   
However, plasma turbulence is well described by ideas that parallel its fluid antecedents, and thus it is instructive to test the hypothesis that coherent structures are linked to nonuniform dissipation using, as a surrogate measure,
the work done by the electromagnetic fields, ${\bf J} \cdot {\bf E}$
in an appropriate reference frame.  

\textit{Analysis}.---We use 450 Hz \textit{burst mode} magnetic field 
$\bf B$ and electric field $\bf E$ 
measurements from the FGM \citep{BaloghEA2001}, 
STAFF \citep{CornilleauWehrlinEA2003} and EFW \citep{GustafssonEA2001} 
instruments onboard the Cluster spacecraft. While components of the dc electric field in the spacecraft spin plane are measured directly, the third component is reconstructed assuming $\mathbf{E}\cdot\mathbf{B} = 0$. We also use 4 s resolution proton moments from the CIS CODIF experiment \citep{RemeEA2001}. 

The curlometer technique \citep{DunlopEA1988, DunlopEA2002} is used to estimate the current density $\mathbf{J}$ through a tetrahedron formed by four spacecraft, where the Maxwell-Ampere law is written as:
\begin{equation}
\mu_{0}\mathbf{J}_{ijk}\cdot(\Delta\mathbf{r}_{ik}\times\Delta\mathbf{r}_{jk}) = \Delta\mathbf{B}_{ik}\cdot\Delta\mathbf{r}_{jk} - \Delta\mathbf{B}_{jk}\cdot\Delta\mathbf{r}_{ik}
\end{equation}
where $i$, $j$ and $k$ are the spacecraft indices, $\mathbf{J}_{ijk}$ is the average current density normal to the surface made by spacecraft $i$, $j$ and $k$, $\Delta\mathbf{r}_{ik} = \mathbf{r}_{i} - \mathbf{r}_{k}$ is the distance between spacecraft $i$ and $k$, and $\Delta\mathbf{B}_{ik} = \mathbf{B}_{i} - \mathbf{B}_{k}$ is the magnetic field difference between spacecraft $i$ and $k$. The total average current density is then determined by projecting the current normal to three faces of the tetrahedron into suitable cartesian coordinates. However, this technique is not without its limitations \citep{VallatEA2005}. The main assumptions are that the spatial variation of 
the magnetic field is a linear function of the spacecraft separation,
such that $\mathbf{J}$ is constant over the tetrahedron,
and that the medium is stationary. 
Since non-stationarity leads to the generation of nonlinear gradients, the only source of error to consider is the nonlinear variation of the magnetic field. This is determined by computing $\nabla\cdot\mathbf{B}$ from:  
\begin{equation}
\nabla\cdot\mathbf{B} | \Delta\mathbf{r}_{ik}\cdot\Delta\mathbf{r}_{jk}\times\Delta\mathbf{r}_{jl} | = | \sum_{cyclic}\Delta\mathbf{B}_{ik}\cdot\Delta\mathbf{r}_{jk}\times\mathbf{r}_{jl} |
\end{equation}
While $\mathbf{B}$ is solenoidal, the expression above can produce non-zero values which result from nonlinear gradients that are neglected in the estimate. Hence, $\nabla\cdot\mathbf{B}/\left|\nabla\times\mathbf{B}\right|$ provides an indicator of the error on curlometer estimates of $\mathbf{J}$. We require this error to be less than 10\%, and remove all data that does not satisfy this condition. Note that certain spacecraft configurations and separations can reduce the accuracy of the curlometer technique, but these effects are likely insignificant in this study since the Cluster quartet are in a regular tetrahedral configuration.

\textit{Results}.---Figure 1 shows selected plasma and magnetic field data for a 12 min interval encompassing an earthward magnetic reconnection jet observed \textit{in-situ} on 17 August 2003 \citep[e.g.][]{HendersonEA2006, AsanoEA2008, HuangEA2012,WangEA2014}. The Cluster quartet are in the magnetotail lobe prior to 16:55 UT, and then enter the plasma sheet whilst located around $[-16.8, 5.6, 3.2]$ Earth radii in geocentric solar magnetospheric (GSM) coordinates with spacecraft separations near 220 km. A high speed earthward flow ($V_{x}$ exceeds 1200 kms$^{-1}$) is detected in the plasma sheet, where the number density $n \approx 0.25$ cm$^{-3}$ and plasma beta $\beta = nk_{B}T/(B^{2}/2\mu_{0}) \approx 1$ have typical values. However, the proton temperature is significantly elevated within the earthward flow, which is suggestive of proton heating by magnetic reconnection. The spacecraft exit the reconnection jet and enter the central plasma sheet at 17:03 UT.

A tailward flow was detected by the spacecraft ahead of the earthward flow from 16:33 to 16:52 UT, and the associated $B_{z}$ was mostly negative (positive) for the tailward (earthward) flow. These correlated changes in $V_{x}$ and $B_{z}$ are consistent with a tailward retreating X-line being swept past the spacecraft, and suggest that the earthward flow is close to this X-line. The negative sign of $B_{y} \approx -15$ nT (roughly 60\% of the asymptotic magnetic field) agrees with the expected Hall magnetic field polarity in the southern hemisphere, eastward of the X-line. Indeed, the reversal in $J_{x}$ observed around 16:55:12 UT could be associated with a reversal in the nearly field-aligned currents which close the Hall currents across the reconnection separatrices. Hence, the spacecraft may have observed an ion diffusion region with a moderate guide field.

\begin{figure}[h]
\includegraphics[width=8.5cm]{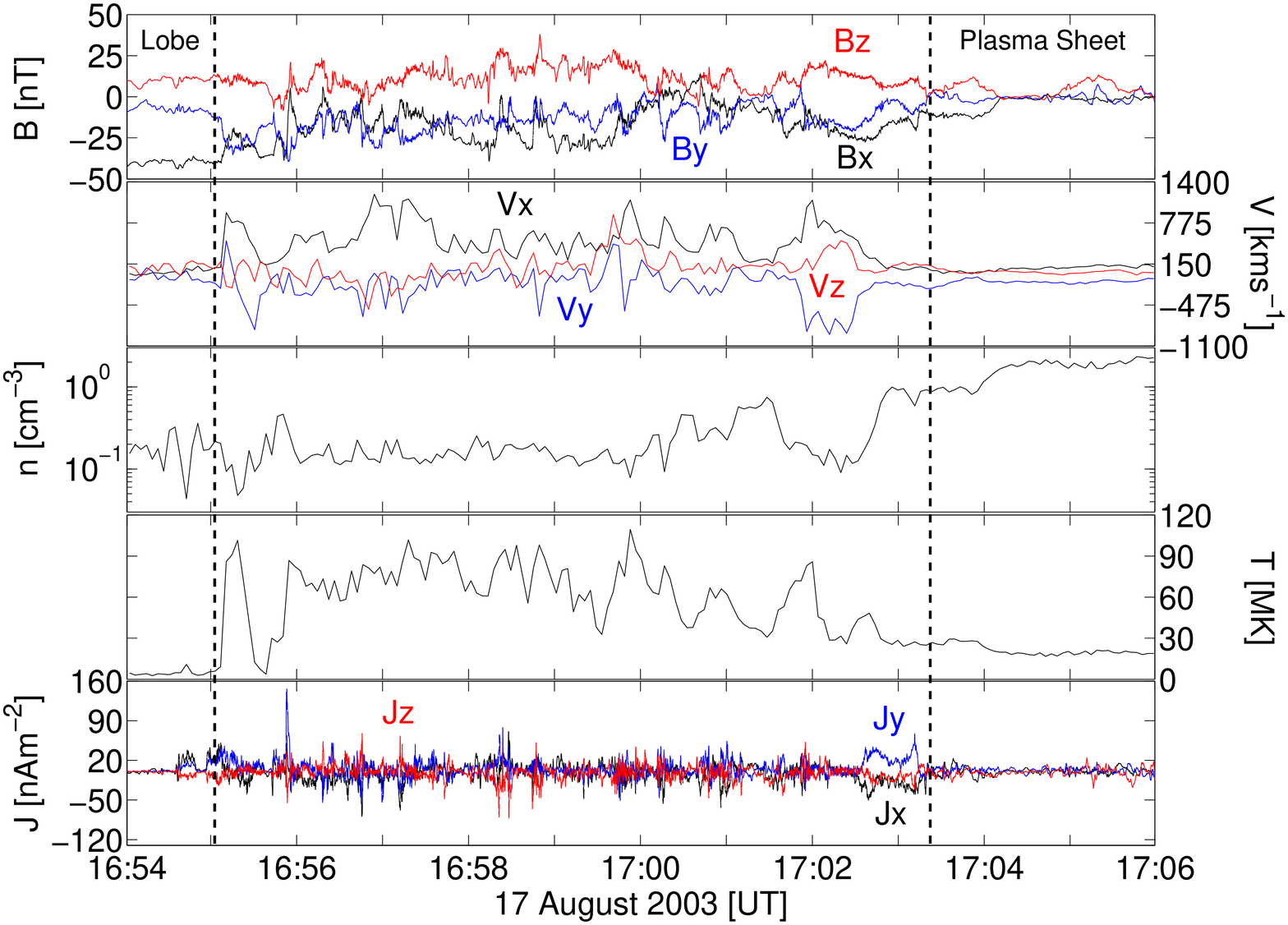}
\caption{An overview of a high speed earthward magnetic reconnection jet. The parameters from top to bottom are: GSM components of magnetic field and solar wind velocity, proton number density, proton temperature and current density. Vertical dashed lines bracket the reconnection jet. The observations before and after the earthward jet are from the magnetotail lobe and central plasma sheet respectively.}
\label{Fig:time}
\end{figure}

There are large fluctuations in the magnetic field and current density associated with the high speed reconnection jet. Here we investigate the statistical properties of these fluctuations to identify signatures of intermittent turbulence. In order to determine the higher-order scaling of magnetic field fluctuations, the absolute moments of the increments $\delta B(t,\tau) = B(t + \tau) - B(t)$ are computed for each vector component
$B \rightarrow B_x, B_y$ or $B_z$. 
The $m$th order structure function is given by: 
\begin{equation}
S^{m}(\tau) = \frac{1}{N}\sum^{N}_{i=1} \left| \delta B(t_{i},\tau) \right|^{m}
\end{equation}
where $\tau$ is the time lag and $N$ is the signal sample size. The higher-order structure functions progressively capture the more intermittent fluctuations. These represent the spatial gradients responsible for dissipating energy in fluid-like turbulence. 
We examine the 
powerlaw 
scaling behavior of structure functions such that:
\begin{equation}
S^{m}(\tau) \propto \tau^{\zeta(m)}
\end{equation}
and $\zeta(m)$ are the scaling exponents. Figure 2 shows the GSM $x$-component magnetic field structure functions and scaling exponents for the earthward flow data interval
indicated in Fig.(\ref{Fig:time}). 
Note that the three 
magnetic field components exhibit essentially 
identical statistical scaling. 
The inertial and dissipation ranges are well defined with a break around 3 s, in agreement with the temporal signature of the proton gyroradius ($\sim$ 260 km). This suggests Taylor's hypothesis \citep{Taylor1938} is valid within the magnetic reconnection jet, and thus each spacecraft time series can be considered a spatial snapshot of the plasma. However, this cannot be confirmed without knowing the characteristic timescale on which the observed fluctuations vary. Note that hereafter we assume time lags and frequency spectra are equivalent to spatial lags and wavenumber spectra. In effect we rely on some form of {\it random sweeping} of small scales \cite{BorovskyEA97}.

The scaling exponent $\zeta(2)$ is directly related to the power spectrum spectral index $\alpha$ when Eq. (4) is satisfied: $\zeta(2) = \alpha - 1$ \citep{MoninYaglom1975}. This relationship only applies to $\alpha < 3$ when using two-point structure functions \citep{KiyaniEA2013}. Here the inertial range follows a power law with $\alpha = 1.65 \pm 0.03$ which then steepens to $\alpha = 2.52 \pm 0.02$ in the dissipation range. These are similar to other reported observations in magnetic reconnection diffusion regions \citep{EastwoodEA2009, HuangEA2010} and in the turbulent solar wind \citep{SmithEA2006}.

\begin{figure}[h]
\includegraphics[width=8.5cm]{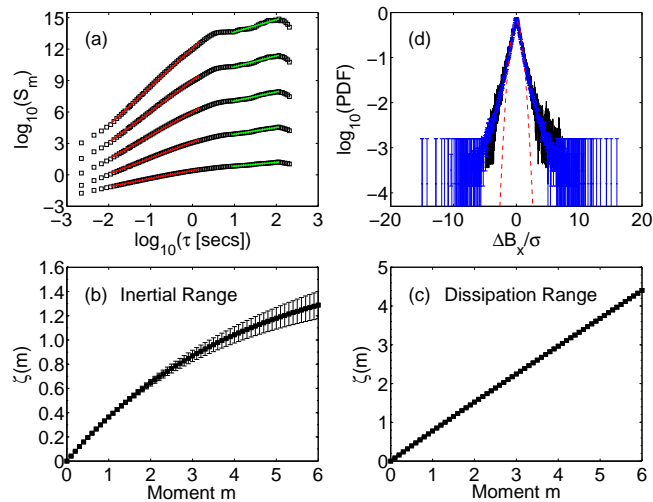}
\caption{The magnetic field (a) structure functions and (b) inertial range scaling exponents. These structure functions of order 1--5 have been shifted along the vertical axis to facilitate comparison of gradients. Linear fits for the inertial and dissipation ranges are also shown. The dissipation range (c) scaling exponents and (d) PDFs rescaled by their standard deviations. A Gaussian fit (dashed curve) is also applied.}
\label{Fig:scaling}
\end{figure}

Figure 2 shows results from the higher-order scaling analysis of magnetic field fluctuations in the reconnection outflow. The errors on $\zeta(m)$ are estimated as the sum of the regression error from Fig. 2(a) and the variation in $\zeta(m)$ found by repeating the regression over a subinterval of the scaling range \citep{KiyaniEA2006}. For inertial range fluctuations, $\zeta(m)$ is nonlinear in $m$,
which is typical of hydrodynamic \citep{Frisch1995} and magnetofluid \citep{Biskamp2003} turbulence. 
This behavior is associated with a statistical 
distribution of energy dissipation that is highly nonuniform, and 
distributed on a spatial multifractal. 
In contrast, dissipation range fluctuations are characterized by a linear $\zeta(m)$. 
This indicates global scale invariance and is associated with 
a distribution of energy dissipation that is also nonuniform, but in this case
distributed on a monofractal. 
Hence, the probability density functions (PDFs) of dissipation range magnetic field increments should collapse onto a unique scaling function. Figure 2(d) shows PDFs corresponding to $\tau = \{0.01, 0.22, 0.44, 0.67, 0.89, 1.11, 1.44\}$ s that are rescaled by their standard deviations and overlaid, where the smallest $\tau$ shows the associated errors. It is apparent that there is a very good
collapse onto a single curve even up to several standard deviations. 
The largest events are not well-sampled as indicated by 
greater statistical spread at larger values of the increments. 
A fitted Gaussian distribution illustrates the non-Gaussian PDF tails, which reflects the presence of rare large amplitude fluctuations. 

These results contradict earlier studies which found multifractal scaling at kinetic scales using PIC simulations \citep{LeonardisEA2013} and from direct observation of the scale-dependent kurtosis \citep{HuangEA2012}. This inconsistency could be because the simulations were set within the complex topology of a reconnection region whereas our observations sample the high speed outflow jet, and that kurtosis is sensitive to very large fluctuations which are not statistically well-sampled in heavy-tailed distributions. However, the higher-order scaling shown in Fig. 2 is almost identical to that observed in the solar wind on both MHD and kinetic scales \citep{KiyaniEA2009, KiyaniEA2013}. This suggests that magnetic field fluctuations generated by reconnection exhibit a detailed correspondence with intermittent turbulence, including a cross-over from multifractal scaling to global scale invariance and distinct non-Gaussian statistics in the inertial and dissipation ranges.

Turbulence cascades energy from structures on larger to smaller spatial scales. The corresponding spatial field is nonuniform with strong fluctuations that have non-Gaussian statistics. These fluctuations are small scale coherent structures which support spatial gradients that can contribute to dissipation \citep{LeonardisEA2013}. 
In analogy with hydrodynamic turbulence this suggests the possibility that reconnection, also related to activity on small scales,
converts magnetic energy into kinetic and thermal energies. 
The connection between these dynamical processes is examined within the earthward flow 
interval by computing 
$D_{l} = \mathbf{J}\cdot\mathbf{E}$, 
the work done by electromagnetic fields on the particles in the spacecraft frame. 
Although this is not strictly a measure of \textit{irreversible} dissipation it must necessarily include the work done to convert stored magnetic energy into heat. Indeed, the identification of $D_{l}$ as dissipation is complicated by contributions from particle acceleration, fluid motion, field line stretching,
and compressions. 
To avoid some of this ambiguity \citep[e.g.][]{ZenitaniEA11}, the work done is evaluated in a frame 
moving with the bulk proton velocity $D_{p} = \mathbf{J}\cdot(\mathbf{E} + \mathbf{V}\times\mathbf{B})$. 

\begin{figure}[h]
\includegraphics[width=8.5cm]{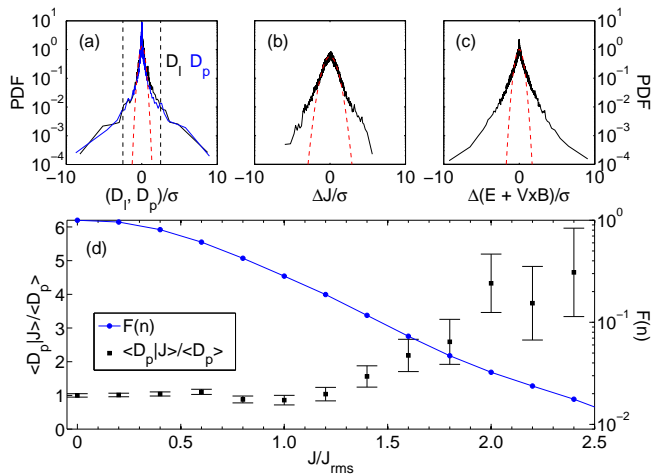}
\caption{PDFs of (a) spacecraft and proton frame work done, (b) current density and (c) electric field fluctuations. Gaussian fits (dashed curves) are applied. The (d) mean proton frame work done conditioned on local current density thresholds, where $J/J_{rms}$ is larger than and equal to some value $n$. Also, the fraction of data in these averages is plotted.}
\label{Fig:dissipation}
\end{figure}

Figure 3(a) shows the broad and slightly asymmetric PDFs of the work done in the laboratory and proton frames rescaled by the standard deviation. These distributions are almost identical, which implies the contribution to the work done from the convective electric field is minimal. There is an excess of positive values in the distribution cores and the average work done is $\langle D_{l} \rangle = 72 \pm 8$ pWm$^{-3}$ and $\langle D_{p} \rangle = 62 \pm 8$ pWm$^{-3}$. These may be interpreted as (imperfect) estimates of net magnetic energy dissipation into plasma internal energy, but will also contain other effects. Note, values beyond the dashed vertical lines ($\pm 2.5\sigma$) are not statistically well-sampled and are likely dominated by unphysical fluctuations on scales smaller than the spacecraft separations. Nonetheless, a fitted Gaussian distribution shows these heavy-tailed PDFs are manifestly non-Gaussian. This is a direct indication of intermittency of dissipation, 
which in the context of the KRSH, is associated with the presence of intermittent fluctuations of the fluid variables such as $\bf B$.
Figures 3(b)-(c) show the current density and electric field fluctuations, which together constitute $D_{p}$, are also independently non-Gaussian and intermittent.

If the work done in the proton frame is highly structured as implied by its leptokurtic PDF, then this inhomogeneity should be evident in suitable conditional statistics. Figure 3(d) shows averages of $D_{p}$ conditioned on thresholds of the 
local current density $\langle D_{p} | J \rangle / \langle D_{p} \rangle$. The fraction of data 
points used in each of 
these conditioned averages is also plotted $F(n) = \sum' f / \sum f$, where $\sum'$ only 
includes points that satisfy the threshold condition $J/J_{rms} \geq n$.
This represents a reasonable and easily accessible measure of the 
volume filling factor of the regions satisfying the corresponding condition. 
These diagnostics illustrate explicitly 
the nonuniform and patchy character of the work done since the normalized 
conditional averages of $D_{p}$ strongly increase with smaller volume fraction. 
This results in a mean $D_{p}$ for the threshold $J \geq 2.4J_{rms}$ that is about 3--6 times the global average, despite such high local current density regions occupying less than 2\% of the data. Hence, regions of higher electric current density are increasingly rare, but make disproportionately large contributions to the total work done.

\textit{Discussion}.---We have used Cluster multispacecraft data to examine the structure within a previously identified magnetic reconnection jet in the terrestrial magnetotail. In particular, the intermittent nature of the magnetic field component increments $\delta B_x$, $\delta B_y$, $\delta B_z$ and the work done, ${\bf J} \cdot {\bf E}$, on particles by the electromagnetic fields is characterized. The significance of this result is seen by recalling the structure of the KRSH for hydrodynamic turbulence: $\delta v_\ell \sim \epsilon_\ell^{1/3} \ell^{1/3}$. It is postulated that the longitudinal velocity increments $\delta v_\ell$ on scale $\ell$ are related statistically to the dissipation rate $\epsilon_l$ averaged over a volume of size $\ell^3$. For a low density plasma such as the magnetospheric plasma sheet, a formal statement of refined similarity has not been elucidated. However, our analysis is tantamount to a statistical examination of both principle elements of a putative analogous relation for plasmas \citep[e.g.][]{MerrifieldEA05,ChandranEA15}, assuming the time increments employed here are comparable to spatial increments which is reasonable for a form of random sweeping. 

The magnetic field fluctuations (increments) within a reconnection jet exhibit a multifractal non-self-similar scaling of higher order moments in the inertial range, which transitions to a self-similar, but still non-Gaussian, monofractal scaling in the the kinetic range. In addition, we find that ${\bf J} \cdot {\bf E}$ within the same jet is highly non-Gaussian, with heavy tails in the probability distribution. Regions of strong ${\bf J} \cdot {\bf E}$ are non-space-filling, with indications that the large transfer of random energy to particles is almost certainly highly concentrated in small volumes that contain atypically large electric current density. Thus, even if the magnetotail plasma is not ohmic in nature, its dissipation is statistically associated with regions of high current density. This finding is consistent with results from the explicit examination of electromagnetic work in two- and three-dimensional collisionless plasma simulations \citep{WanEA12,KarimabadiEA13,WanEA15}, and from less direct inference in observations of the solar wind \citep{OsmanEA2014}, magnetopause \citep{MozerEA2002} and magnetosheath \citep{SundkvistEA07}. The conclusion seems increasingly certain that intermittent dissipation is as typical in large low density plasma systems as it is in high Reynolds number hydrodynamic turbulence.

While our analysis focuses on turbulence within a magnetic reconnection outflow jet, the reconnection process itself occurs within a small diffusion region. The NASA Magnetospheric Multi-Scale mission will make high-resolution plasma and magnetic field measurements, and thus should allow the statistical nature of ${\bf J} \cdot {\bf E}$ to be probed within the proton and electron diffusion regions. This will improve our understanding of magnetic reconnection and turbulent dissipation. Indeed, as diagnostics for examining intermittent dissipation become more well understood, we may anticipate a more refined expectation may be emerging regarding the heating and intermittency that will likely be observed by the upcoming ESA Solar Orbiter, NASA Solar Probe Plus and proposed ESA THOR missions as they seek to understand how the solar corona is heated and the solar wind is accelerated, leading to the emerging structure of the entire plasma heliosphere.

This research is supported by the UK STFC under grant ST/I000720/1, NASA through the the MMS theory and Modeling team (NNX14AC39G) and the Heliospheric Grand Challenge Research program (NNX14AI63G), and
by the NSF (AGS-1063439, SHINE AGS-1156094).


\begin{thebibliography}{99}

\bibitem{ZimbardoEA2010}
G. Zimbardo et al., Space Sci. Rev. 156, 89 (2010).

\bibitem{AnselmetEA1984}
F. Anselmet, Y. Gagne, E. J. Hopfinger, \& R. A. Antonia, J. Fluid Mech. 140, 63 (1984).

\bibitem{MatthaeusMontgomery1980}
W. H. Matthaeus \& D. Montgomery, Ann, N. Y. Acad. Sci. 357, 203 (1980).

\bibitem{KarimabadiEA13}
H. Karimabadi et al., Phys. Plasmas 20, 012303 (2013).

\bibitem{OsmanEA2014}
K. T. Osman et al., Phys. Rev. Lett. 112, 215002 (2014).

\bibitem{MozerEA2002}
F. S. Mozer, S. D. Bale, \& T. D. Phan, Phys. Rev. Lett. 89, 015002 (2002).

\bibitem{RetinoEA2007}
A. Retin\'o et al., Nature Phys. 3, 236 (2007).

\bibitem{SundkvistEA07}
D. Sundkvist, A. Retin\'o, A. Vaivads and S. D. Bale, Phys. Rev. Lett. 99, 025004 (2007).

\bibitem{VorosEA2004}
Z. V\"or\"os et al., J. Geophys. Res. 109, A11215 (2004).

\bibitem{WeygandEA2005}
J. M. Weygand et al., J. Geophys. Res. 110, A01205 (2005).

\bibitem{BorovskyEA1997}
J. E. Borovsky, R. C. Elphic, H. O. Funsten, \& M. F. Thomsen, J. Plasma Phys. 57, 1 (1997).

\bibitem{BorovskyFunsten2003}
J. E. Borovsky \& H. O. Funsten, J. Geophys. Res. 108, A009625 (2003)

\bibitem{Chang1999}
T. Chang, Phys. Plasmas 6, 4137 (1999).

\bibitem{MatthaeusLamkin1986}
W. H. Matthaeus \& S. L. Lamkin, Phys. Fluids 29, 2513 (1986).

\bibitem{MatthaeusEA2014}
W. H. Matthaeus et al., Astrophys. J. 790, 155 (2014).

\bibitem{Frisch1995}
U. Frisch, Turbulence, Cambridge: Cambridge Univ. Press (1995).

\bibitem{Biskamp2003}
D. Biskamp, Magnetohydrodynamic Turbulence, Cambridge: Cambridge University Press (2003).

\bibitem{Kolmogorov1962}
A.N. Kolmogorov, J. Fluid Mech. 12, 82 (1962).

\bibitem{Obukhov1962}
A.M. Obukhov, J. Fluid Mech. 13, 77 (1962).

\bibitem{SreenivasanAntonia1997}
K. R. Sreenivasan \& R. A. Antonia, Ann. Rev. Fluid Mech. 29, 435 (1997).

\bibitem{BaloghEA2001}
A. Balogh et al., Ann. Geophys. 19, 1207 (2001).

\bibitem{CornilleauWehrlinEA2003}
N. Cornilleau-Wehrlin et al., Ann. Geophys. 21, 437 (2003).

\bibitem{GustafssonEA2001}
G. Gustafsson et al., Ann. Geophys. 19, 1219 (2001).

\bibitem{RemeEA2001}
H. R\'eme et al., Ann. Geophys. 19, 1303 (2001). 

\bibitem{DunlopEA1988}
M. W. Dunlop, D. J. Southwood, K. -H. Glassmeier, \& F. M. Neubauer, Adv. Space Res. 8, 9 (1988).

\bibitem{DunlopEA2002}
M. W. Dunlop, A. Balogh, K. -H. Glassmeier, \& P. Robert, J. Geophys. Res. 107, 1384 (2002).

\bibitem{VallatEA2005}
C. Vallat et al., Ann. Geophys. 23, 1849 (2005).

\bibitem{HendersonEA2006}
P. D. Henderson et al., Geophys. Res. Lett. 33, L22106 (2006).

\bibitem{AsanoEA2008}
Y. Asano et al., J. Geophys. Res. 113, A01207 (2008)

\bibitem{HuangEA2012}
S. Y. Huang et al., Geophys. Res. Lett. 39, L11104 (2012).

\bibitem{WangEA2014}
R. Wang et al., Geophys. Res. Lett. 41, 4851 (2014).

\bibitem{Taylor1938}
G. I. Taylor, Proc. R. Soc. A 164, 476 (1938).

\bibitem{BorovskyEA97}
J. E. Borovsky, R. C. Elphic, H. O. Funsten, \& M. F. Thomsen, Journal of Plasma Physics 57, 1 (1997).

\bibitem{MoninYaglom1975}
A. Monin \& A. Yaglom, Statistical Fluid Mechanics: Mechanics of Turbulence, vol. 2, M.I.T Press, Cambridge, Massachusetts (1975).

\bibitem{EastwoodEA2009}
J. P. Eastwood, T. D. Phan, S. D. Bale, \& A. Tjulin, Phys. Rev. Lett. 102, 035001 (2009).

\bibitem{HuangEA2010}
S. Y. Huang et al., J. Geophys. Res. 115, A12211 (2010).

\bibitem{SmithEA2006}
C. W. Smith, K. Hamilton, B. J. Vasquez, \& R. J. Leamon, Astrophys. J. 645, L85 (2006).

\bibitem{KiyaniEA2006}
K. H. Kiyani, S. C. Chapman, \& B. Hnat, Phys. Rev. E 74, 051122 (2006).

\bibitem{LeonardisEA2013}
E. Leonardis, S. C. Chapman, W. Daughton, V. Roytershteyn, \& H. Karimabadi, Phys. Rev. Lett. 110, 205002 (2013).

\bibitem{KiyaniEA2009}
K. H. Kiyani, S. C. Chapman, Yu. V. Khotyaintsev, M. W. Dunlop, \& F. Sahraoui, Phys. Rev. Lett. 103, 075006 (2009).

\bibitem{KiyaniEA2013}
K. H. Kiyani et al., Astrophys. J. 763, 10 (2013). 

\bibitem{MerrifieldEA05}
J. A. Merrifield, W. C. M{\"u}ller, S. C. Chapman, \& R. O. Dendy, Physics of Plasmas 12, 022301 (2005).

\bibitem{ChandranEA15} 
B. D. G. Chandran, A. A. Schekochihin, \& A. Mallet, Astrophys. J. 807, 39 (2015).

\bibitem{WanEA12}
M. Wan et al., Phys. Rev. Lett. 109, 195001 (2012).

\bibitem{WanEA15} 
M. Wan, W. H. Matthaeus, V. Roytershteyn, H. Karimabadi, T. Parashar, P. Wu \& M. Shay, Phys. Rev. Lett. 114, 175002 (2015).

\bibitem{ChasapisEA15} 
A. Chasapis et al., Astrophys. J. 804, L1 (2015). 

\bibitem{ZenitaniEA11}
S. Zenitani, M. Hesse, A. Klimas, \& M. Kuznetsova, Phys. Rev. Lett. 106, 195003 (2011).

\end{thebibliography}
\end{document}